\documentclass[journal]{IEEEtran}
\usepackage{xcolor}
\usepackage{cite}
\usepackage{amsmath}
\usepackage{amsfonts}
\usepackage[short]{optidef}
\usepackage{subfig}
\usepackage[pdftex]{graphicx}
\usepackage{autobreak}
\usepackage{comment}
\usepackage{url}
\usepackage{booktabs}
\usepackage{amsmath}
\usepackage[acronyms,nonumberlist,nopostdot,nomain,nogroupskip,acronymlists={hidden}]{glossaries}
\newglossary[algh]{hidden}{acrh}{acnh}{Hidden Acronyms}
\newacronym{3gpp}{3GPP}{3rd Generation Partnership Project}
\newacronym{4g}{4G}{4th generation}
\newacronym{5g}{5G}{5th generation}
\newacronym{6g}{6G}{6th generation}
\newacronym{5gc}{5GC}{5G Core}
\newacronym{adc}{ADC}{Analog to Digital Converter}
\newacronym{aerpaw}{AERPAW}{Aerial Experimentation and Research Platform for Advanced Wireless}
\newacronym{ai}{AI}{Artificial Intelligence}
\newacronym{EM}{EM}{electromagnetic}
\newacronym{EMS}{EMS}{electromagnetic skin}
\newacronym{mmW}{mmW}{millimeter wave}
\newacronym{RoI}{RoI}{region of interest}
\newacronym{RIS}{RIS}{Reconfigurable Intelligent Surface}
\newacronym{RISs}{RISs}{Reconfigurable Intelligent Surfaces}
\newacronym{SNR}{SNR}{signal-to-noise ratio}
\newacronym{AF}{AF}{amplify-and-forward}
\newacronym{DF}{DF}{decode-and-forward}
\newacronym{STAR-RISs}{STAR-RISs}{Simultaneous Transmit and Reflecting RISs}
\newacronym{3GPP}{3GPP}{3rd Generation Partnership Project}
\newacronym{RAN}{RAN}{Radio Access Network}
\newacronym{NCR}{NCR}{Network-Controlled Repeater}
\newacronym{NCRs}{NCRs}{Network-Controlled Repeaters}
\newacronym{IAB}{IAB}{Integated-Access-and-Backhauling}
\newacronym{SRE}{SRE}{Smart Radio Environment}
\newacronym{HSRE}{HSRE}{Heterogeneous SRE}
\newacronym{aimd}{AIMD}{Additive Increase Multiplicative Decrease}
\newacronym{am}{AM}{Acknowledged Mode}
\newacronym{amc}{AMC}{Adaptive Modulation and Coding}
\newacronym{amf}{AMF}{Access and Mobility Management Function}
\newacronym{aops}{AOPS}{Adaptive Order Prediction Scheduling}
\newacronym{api}{API}{Application Programming Interface}
\newacronym{apn}{APN}{Access Point Name}
\newacronym{ap}{AP}{Application Protocol}
\newacronym{aqm}{AQM}{Active Queue Management}
\newacronym{ar}{AR}{Augmented Reality}
\newacronym{ausf}{AUSF}{Authentication Server Function}
\newacronym{avc}{AVC}{Advanced Video Coding}
\newacronym{awgn}{AGWN}{Additive White Gaussian Noise}
\newacronym{balia}{BALIA}{Balanced Link Adaptation Algorithm}
\newacronym{bbu}{BBU}{Base Band Unit}
\newacronym{bdp}{BDP}{Bandwidth-Delay Product}
\newacronym{ber}{BER}{Bit Error Rate}
\newacronym{bf}{BF}{Beamforming}
\newacronym{bler}{BLER}{Block Error Rate}
\newacronym{brr}{BRR}{Bayesian Ridge Regressor}
\newacronym{BS}{BS}{Base Station}
\newacronym{bsr}{BSR}{Buffer Status Report}
\newacronym{bss}{BSS}{Business Support System}
\newacronym{ca}{CA}{Carrier Aggregation}
\newacronym{caas}{CaaS}{Connectivity-as-a-Service}
\newacronym{cb}{CB}{Code Block}
\newacronym{cc}{CC}{Congestion Control}
\newacronym{ccid}{CCID}{Congestion Control ID}
\newacronym{cco}{CC}{Carrier Component}
\newacronym{cdd}{CDD}{Cyclic Delay Diversity}
\newacronym{cdf}{CDF}{Cumulative Distribution Function}
\newacronym{cdn}{CDN}{Content Distribution Network}
\newacronym{cn}{CN}{Core Network}
\newacronym{codel}{CoDel}{Controlled Delay Management}
\newacronym{comac}{COMAC}{Converged Multi-Access and Core}
\newacronym{cord}{CORD}{Central Office Re-architected as a Datacenter}
\newacronym{cornet}{CORNET}{COgnitive Radio NETwork}
\newacronym{cosmos}{COSMOS}{Cloud Enhanced Open Software Defined Mobile Wireless Testbed for City-Scale Deployment}
\newacronym{cots}{COTS}{Commercial Off-the-Shelf}
\newacronym{cp}{CP}{Control Plane}
\newacronym{cyp}{CP}{Cyclic Prefix}
\newacronym{up}{UP}{User Plane}
\newacronym{cpu}{CPU}{Central Processing Unit}
\newacronym{cqi}{CQI}{Channel Quality Information}
\newacronym{cr}{CR}{Cognitive Radio}
\newacronym{cran}{C-RAN}{Cloud \gls{ran}}
\newacronym{crs}{CRS}{Cell Reference Signal}
\newacronym{csi}{CSI}{Channel State Information}
\newacronym{csirs}{CSI-RS}{Channel State Information - Reference Signal}
\newacronym{cu}{CU}{Central Unit}
\newacronym{d2tcp}{D$^2$TCP}{Deadline-aware Data center TCP}
\newacronym{d3}{D$^3$}{Deadline-Driven Delivery}
\newacronym{dac}{DAC}{Digital to Analog Converter}
\newacronym{dag}{DAG}{Directed Acyclic Graph}
\newacronym{das}{DAS}{Distributed Antenna System}
\newacronym{dash}{DASH}{Dynamic Adaptive Streaming over HTTP}
\newacronym{dc}{DC}{Dual Connectivity}
\newacronym{dccp}{DCCP}{Datagram Congestion Control Protocol}
\newacronym{dce}{DCE}{Direct Code Execution}
\newacronym{dci}{DCI}{Downlink Control Information}
\newacronym{dctcp}{DCTCP}{Data Center TCP}
\newacronym{dl}{DL}{Downlink}
\newacronym{dmr}{DMR}{Deadline Miss Ratio}
\newacronym{dmrs}{DMRS}{DeModulation Reference Signal}
\newacronym{drlcc}{DRL-CC}{Deep Reinforcement Learning Congestion Control}
\newacronym{drs}{DRS}{Discovery Reference Signal}
\newacronym{du}{DU}{Distributed Unit}
\newacronym{e2e}{E2E}{end-to-end}
\newacronym{ecaas}{ECaaS}{Edge-Cloud-as-a-Service}
\newacronym{ecn}{ECN}{Explicit Congestion Notification}
\newacronym{edf}{EDF}{Earliest Deadline First}
\newacronym{embb}{eMBB}{Enhanced Mobile Broadband}
\newacronym{empower}{EMPOWER}{EMpowering transatlantic PlatfOrms for advanced WirEless Research}
\newacronym{enb}{eNB}{evolved Node Base}
\newacronym{endc}{EN-DC}{E-UTRAN-\gls{nr} \gls{dc}}
\newacronym{epc}{EPC}{Evolved Packet Core}
\newacronym{eps}{EPS}{Evolved Packet System}
\newacronym{es}{ES}{Edge Server}
\newacronym{etsi}{ETSI}{European Telecommunications Standards Institute}
\newacronym[firstplural=Estimated Times of Arrival (ETAs)]{eta}{ETA}{Estimated Time of Arrival}
\newacronym{eutran}{E-UTRAN}{Evolved Universal Terrestrial Access Network}
\newacronym{faas}{FaaS}{Function-as-a-Service}
\newacronym{fapi}{FAPI}{Functional Application Platform Interface}
\newacronym{fdd}{FDD}{Frequency Division Duplexing}
\newacronym{fdm}{FDM}{Frequency Division Multiplexing}
\newacronym{fdma}{FDMA}{Frequency Division Multiple Access}
\newacronym{fed4fire}{FED4FIRE+}{Federation 4 Future Internet Research and Experimentation Plus}
\newacronym{fir}{FIR}{Finite Impulse Response}
\newacronym{fit}{FIT}{Future \acrlong{iot}}
\newacronym{fpga}{FPGA}{Field Programmable Gate Array}
\newacronym{fr2}{FR2}{Frequency Range 2}
\newacronym{fs}{FS}{Fast Switching}
\newacronym{fscc}{FSCC}{Flow Sharing Congestion Control}
\newacronym{ftp}{FTP}{File Transfer Protocol}
\newacronym{fw}{FW}{Flow Window}
\newacronym{ge}{GE}{Gaussian Elimination}
\newacronym{gnb}{gNB}{Next Generation Node Base}
\newacronym{gop}{GOP}{Group of Pictures}
\newacronym{gpr}{GPR}{Gaussian Process Regressor}
\newacronym{gpu}{GPU}{Graphics Processing Unit}
\newacronym{gtp}{GTP}{GPRS Tunneling Protocol}
\newacronym{gtpc}{GTP-C}{GPRS Tunnelling Protocol Control Plane}
\newacronym{gtpu}{GTP-U}{GPRS Tunnelling Protocol User Plane}
\newacronym{gtpv2c}{GTPv2-C}{\gls{gtp} v2 - Control}
\newacronym{gw}{GW}{Gateway}
\newacronym{harq}{HARQ}{Hybrid Automatic Repeat reQuest}
\newacronym{hetnet}{HetNet}{Heterogeneous Network}
\newacronym{hh}{HH}{Hard Handover}
\newacronym{hol}{HOL}{Head-of-Line}
\newacronym{hqf}{HQF}{Highest-quality-first}
\newacronym{hss}{HSS}{Home Subscription Server}
\newacronym{http}{HTTP}{HyperText Transfer Protocol}
\newacronym{ia}{IA}{Initial Access}
\newacronym{iab}{IAB}{Integrated Access and Backhaul}
\newacronym{ic}{IC}{Incident Command}
\newacronym{ietf}{IETF}{Internet Engineering Task Force}
\newacronym{imsi}{IMSI}{International Mobile Subscriber Identity}
\newacronym{imt}{IMT}{International Mobile Telecommunication}
\newacronym{iot}{IoT}{Internet of Things}
\newacronym{ip}{IP}{Internet Protocol}
\newacronym{itu}{ITU}{International Telecommunication Union}
\newacronym{kpi}{KPI}{Key Performance Indicator}
\newacronym{kpm}{KPM}{Key Performance Measurement}
\newacronym{kvm}{KVM}{Kernel-based Virtual Machine}
\newacronym{los}{LOS}{Line-of-Sight}
\newacronym{lsm}{LSM}{Link-to-System Mapping}
\newacronym{lstm}{LSTM}{Long Short Term Memory}
\newacronym{lte}{LTE}{Long Term Evolution}
\newacronym{lxc}{LXC}{Linux Container}
\newacronym{m2m}{M2M}{Machine to Machine}
\newacronym{mac}{MAC}{Medium Access Control}
\newacronym{manet}{MANET}{Mobile Ad Hoc Network}
\newacronym{mano}{MANO}{Management and Orchestration}
\newacronym{mc}{MC}{Multi-Connectivity}
\newacronym{mcc}{MCC}{Mobile Cloud Computing}
\newacronym{mchem}{MCHEM}{Massive Channel Emulator}
\newacronym{mcs}{MCS}{Modulation and Coding Scheme}
\newacronym{mec}{MEC}{Multi-access Edge Computing}
\newacronym{mec2}{MEC}{Mobile Edge Cloud}
\newacronym{mfc}{MFC}{Mobile Fog Computing}
\newacronym{mgen}{MGEN}{Multi-Generator}
\newacronym{mi}{MI}{Mutual Information}
\newacronym{mib}{MIB}{Master Information Block}
\newacronym{miesm}{MIESM}{Mutual Information Based Effective SINR}
\newacronym{mimo}{MIMO}{Multiple Input, Multiple Output}
\newacronym{ml}{ML}{Machine Learning}
\newacronym{mlr}{MLR}{Maximum-local-rate}
\newacronym[plural=\gls{mme}s,firstplural=Mobility Management Entities (MMEs)]{mme}{MME}{Mobility Management Entity}
\newacronym{mmtc}{mMTC}{Massive Machine-Type Communications}
\newacronym{mmwave}{mmWave}{millimeter wave}
\newacronym{mpdccp}{MP-DCCP}{Multipath Datagram Congestion Control Protocol}
\newacronym{mptcp}{MPTCP}{Multipath TCP}
\newacronym{mr}{MR}{Maximum Rate}
\newacronym{mrdc}{MR-DC}{Multi \gls{rat} \gls{dc}}
\newacronym{mse}{MSE}{Mean Square Error}
\newacronym{mss}{MSS}{Maximum Segment Size}
\newacronym{mt}{MT}{Mobile Termination}
\newacronym{mtd}{MTD}{Machine-Type Device}
\newacronym{mtu}{MTU}{Maximum Transmission Unit}
\newacronym{mumimo}{MU-MIMO}{Multi-user \gls{mimo}}
\newacronym{mvno}{MVNO}{Mobile Virtual Network Operator}
\newacronym{nalu}{NALU}{Network Abstraction Layer Unit}
\newacronym{nas}{NAS}{Non-Access Stratum}
\newacronym{nbiot}{NB-IoT}{Narrow Band IoT}
\newacronym{nfv}{NFV}{Network Function Virtualization}
\newacronym{nfvi}{NFVI}{Network Function Virtualization Infrastructure}
\newacronym{ngrg}{nGRG}{next Generation Research Group}
\newacronym{ni}{NI}{Network Interfaces}
\newacronym{nic}{NIC}{Network Interface Card}
\newacronym{nlos}{NLOS}{Non-Line-of-Sight}
\newacronym{now}{NOW}{Non Overlapping Window}
\newacronym{nsm}{NSM}{Network Service Mesh}
\newacronym{nr}{NR}{New Radio}
\newacronym{nrf}{NRF}{Network Repository Function}
\newacronym{nsa}{NSA}{Non Stand Alone}
\newacronym{nse}{NSE}{Network Slicing Engine}
\newacronym{nssf}{NSSF}{Network Slice Selection Function}
\newacronym{o2i}{O2I}{Outdoor to Indoor}
\newacronym{oai}{OAI}{OpenAirInterface}
\newacronym{oaicn}{OAI-CN}{\gls{oai} \acrlong{cn}}
\newacronym{oairan}{OAI-RAN}{\acrlong{oai} \acrlong{ran}}
\newacronym{oam}{OAM}{Operations, Administration and Maintenance}
\newacronym{ofdm}{OFDM}{Orthogonal Frequency Division Multiplexing}
\newacronym{olia}{OLIA}{Opportunistic Linked Increase Algorithm}
\newacronym{omec}{OMEC}{Open Mobile Evolved Core}
\newacronym{onap}{ONAP}{Open Network Automation Platform}
\newacronym{onf}{ONF}{Open Networking Foundation}
\newacronym{onos}{ONOS}{Open Networking Operating System}
\newacronym{oom}{OOM}{\gls{onap} Operations Manager}
\newacronym{opnfv}{OPNFV}{Open Platform for \gls{nfv}}
\newacronym{oran}{O-RAN}{Open Radio Access Network}
\newacronym{orbit}{ORBIT}{Open-Access Research Testbed for Next-Generation Wireless Networks}
\newacronym{os}{OS}{Operating System}
\newacronym{oss}{OSS}{Operations Support System}
\newacronym{otic}{OTIC}{Open Testing \& Integration Centre}
\newacronym{pa}{PA}{Position-aware}
\newacronym{pase}{PASE}{Prioritization, Arbitration, and Self-adjusting Endpoints}
\newacronym{pawr}{PAWR}{Platforms for Advanced Wireless Research}
\newacronym{pbch}{PBCH}{Physical Broadcast Channel}
\newacronym{pcef}{PCEF}{Policy and Charging Enforcement Function}
\newacronym{pcfich}{PCFICH}{Physical Control Format Indicator Channel}
\newacronym{pcrf}{PCRF}{Policy and Charging Rules Function}
\newacronym{pdcch}{PDCCH}{Physical Downlink Control Channel}
\newacronym{pdcp}{PDCP}{Packet Data Convergence Protocol}
\newacronym{pdsch}{PDSCH}{Physical Downlink Shared Channel}
\newacronym{pdu}{PDU}{Packet Data Unit}
\newacronym{pf}{PF}{Proportional Fair}
\newacronym{pgw}{PGW}{Packet Gateway}
\newacronym{phich}{PHICH}{Physical Hybrid ARQ Indicator Channel}
\newacronym{phy}{PHY}{Physical}
\newacronym{pmch}{PMCH}{Physical Multicast Channel}
\newacronym{pmi}{PMI}{Precoding Matrix Indicators}
\newacronym{powder}{POWDER}{Platform for Open Wireless Data-driven Experimental Research}
\newacronym{ppo}{PPO}{Proximal Policy Optimization}
\newacronym{ppp}{PPP}{Poisson Point Process}
\newacronym{prach}{PRACH}{Physical Random Access Channel}
\newacronym{prb}{PRB}{Physical Resource Block}
\newacronym{psnr}{PSNR}{Peak Signal to Noise Ratio}
\newacronym{pss}{PSS}{Primary Synchronization Signal}
\newacronym{pucch}{PUCCH}{Physical Uplink Control Channel}
\newacronym{pusch}{PUSCH}{Physical Uplink Shared Channel}
\newacronym{rar}{RAR}{Random Access Response}
\newacronym{qam}{QAM}{Quadrature Amplitude Modulation}
\newacronym{qci}{QCI}{\gls{qos} Class Identifier}
\newacronym{5qi}{5QI}{5G \gls{qos} Identifier}
\newacronym{qoe}{QoE}{Quality of Experience}
\newacronym{QoS}{QoS}{Quality of Service}
\newacronym{UE}{UE}{User Equipment}
\newacronym{UEs}{UEs}{User Equipments}
\newacronym{FoV}{FoV}{field of view}
\newacronym{UPA}{UPA}{uniform planar array}
\newacronym{quic}{QUIC}{Quick UDP Internet Connections}
\newacronym{rach}{RACH}{Random Access Channel}
\newacronym{ran}{RAN}{Radio Access Network}
\newacronym[firstplural=Radio Access Technologies (RATs)]{rat}{RAT}{Radio Access Technology}
\newacronym{rcn}{RCN}{Research Coordination Network}
\newacronym{STAR}{STAR-RIS}{simultaneous transmitting and reflecting RIS}
\newacronym{3SNCR}{3SNCR}{trisectoral NCR}
\newacronym{rc}{RC}{RAN Control}
\newacronym{rec}{REC}{Radio Edge Cloud}
\newacronym{red}{RED}{Random Early Detection}
\newacronym{renew}{RENEW}{Reconfigurable Eco-system for Next-generation End-to-end Wireless}
\newacronym{rf}{RF}{Radio Frequency}
\newacronym{rfc}{RFC}{Request for Comments}
\newacronym{rfr}{RFR}{Random Forest Regressor}
\newacronym{ric}{RIC}{\gls{ran} Intelligent Controller}
\newacronym{rlc}{RLC}{Radio Link Control}
\newacronym{rlf}{RLF}{Radio Link Failure}
\newacronym{rlnc}{RLNC}{Random Linear Network Coding}
\newacronym{rmr}{RMR}{RIC Message Router}
\newacronym{rmse}{RMSE}{Root Mean Squared Error}
\newacronym{rnis}{RNIS}{Radio Network Information Service}
\newacronym{rr}{RR}{Round Robin}
\newacronym{rrc}{RRC}{Radio Resource Control}
\newacronym{rrm}{RRM}{Radio Resource Management}
\newacronym{rru}{RRU}{Remote Radio Unit}
\newacronym{rs}{RS}{Remote Server}
\newacronym{rsrp}{RSRP}{Reference Signal Received Power}
\newacronym{rsrq}{RSRQ}{Reference Signal Received Quality}
\newacronym{rss}{RSS}{Received Signal Strength}
\newacronym{rssi}{RSSI}{Received Signal Strength Indicator}
\newacronym{rtt}{RTT}{Round Trip Time}
\newacronym{ru}{RU}{Radio Unit}
\newacronym{rw}{RW}{Receive Window}
\newacronym{rx}{RX}{Receiver}
\newacronym{s1ap}{S1AP}{S1 Application Protocol}
\newacronym{sa}{SA}{standalone}
\newacronym{sack}{SACK}{Selective Acknowledgment}
\newacronym{sap}{SAP}{Service Access Point}
\newacronym{sc2}{SC2}{Spectrum Collaboration Challenge}
\newacronym{scef}{SCEF}{Service Capability Exposure Function}
\newacronym{sch}{SCH}{Secondary Cell Handover}
\newacronym{scoot}{SCOOT}{Split Cycle Offset Optimization Technique}
\newacronym{sctp}{SCTP}{Stream Control Transmission Protocol}
\newacronym{sdap}{SDAP}{Service Data Adaptation Protocol}
\newacronym{sdk}{SDK}{Software Development Kit}
\newacronym{sdm}{SDM}{Space Division Multiplexing}
\newacronym{sdma}{SDMA}{Spatial Division Multiple Access}
\newacronym{sdn}{SDN}{Software-defined Networking}
\newacronym{sdr}{SDR}{Software-defined Radio}
\newacronym{seba}{SEBA}{SDN-Enabled Broadband Access}
\newacronym{sgsn}{SGSN}{Serving GPRS Support Node}
\newacronym{sgw}{SGW}{Service Gateway}
\newacronym{si}{SI}{Study Item}
\newacronym{sib}{SIB}{Secondary Information Block}
\newacronym{sinr}{SINR}{Signal to Interference plus Noise Ratio}
\newacronym{sip}{SIP}{Session Initiation Protocol}
\newacronym{siso}{SISO}{Single Input, Single Output}
\newacronym{sla}{SLA}{Service Level Agreement}
\newacronym{sm}{SM}{Service Model}
\newacronym{smf}{SMF}{Session Management Function}
\newacronym{smo}{SMO}{Service Management and Orchestration}
\newacronym{sms}{SMS}{Short Message Service}
\newacronym{smsgmsc}{SMS-GMSC}{\gls{sms}-Gateway}
\newacronym{snr}{SNR}{Signal-to-Noise-Ratio}
\newacronym{son}{SON}{Self-Organizing Network}
\newacronym{sptcp}{SPTCP}{Single Path TCP}
\newacronym{srb}{SRB}{Service Radio Bearer}
\newacronym{srn}{SRN}{Standard Radio Node}
\newacronym{srs}{SRS}{Sounding Reference Signal}
\newacronym{zc}{ZC}{Zadoff-Chu}
\newacronym{ta}{TA}{Timing Advance}
\newacronym{ss}{SS}{Synchronization Signal}
\newacronym{sss}{SSS}{Secondary Synchronization Signal}
\newacronym{st}{ST}{Spanning Tree}
\newacronym{svc}{SVC}{Scalable Video Coding}
\newacronym{tb}{TB}{Transport Block}
\newacronym{tcp}{TCP}{Transmission Control Protocol}
\newacronym{tdd}{TDD}{Time Division Duplexing}
\newacronym{tdm}{TDM}{Time Division Multiplexing}
\newacronym{tdma}{TDMA}{Time Division Multiple Access}
\newacronym{tfl}{TfL}{Transport for London}
\newacronym{tfrc}{TFRC}{TCP-Friendly Rate Control}
\newacronym{tft}{TFT}{Traffic Flow Template}
\newacronym{tgen}{TGEN}{Traffic Generator}
\newacronym{tip}{TIP}{Telecom Infra Project}
\newacronym{tm}{TM}{Transparent Mode}
\newacronym{to}{TO}{Telco Operator}
\newacronym{tr}{TR}{Technical Report}
\newacronym{trp}{TRP}{Transmitter Receiver Pair}
\newacronym{ts}{TS}{Technical Specification}
\newacronym{tti}{TTI}{Transmission Time Interval}
\newacronym{ttt}{TTT}{Time-to-Trigger}
\newacronym{tx}{TX}{Transmitter}
\newacronym{uas}{UAS}{Unmanned Aerial System}
\newacronym{uav}{UAV}{Unmanned Aerial Vehicle}
\newacronym{udm}{UDM}{Unified Data Management}
\newacronym{udp}{UDP}{User Datagram Protocol}
\newacronym{udr}{UDR}{Unified Data Repository}
\newacronym{ue}{UE}{User Equipment}
\newacronym{uhd}{UHD}{\gls{usrp} Hardware Driver}
\newacronym{ul}{UL}{Uplink}
\newacronym{um}{UM}{Unacknowledged Mode}
\newacronym{uml}{UML}{Unified Modeling Language}
\newacronym{upa}{UPA}{Uniform Planar Array}
\newacronym{upf}{UPF}{User Plane Function}
\newacronym{urllc}{URLLC}{Ultra Reliable and Low Latency Communications}
\newacronym{usa}{U.S.}{United States}
\newacronym{usim}{USIM}{Universal Subscriber Identity Module}
\newacronym{usrp}{USRP}{Universal Software Radio Peripheral}
\newacronym{utc}{UTC}{Urban Traffic Control}
\newacronym{vim}{VIM}{Virtualization Infrastructure Manager}
\newacronym{vm}{VM}{Virtual Machine}
\newacronym{vnf}{VNF}{Virtual Network Function}
\newacronym{volte}{VoLTE}{Voice over \gls{lte}}
\newacronym{voltha}{VOLTHA}{Virtual OLT HArdware Abstraction}
\newacronym{vr}{VR}{Virtual Reality}
\newacronym{vran}{vRAN}{Virtualized \gls{ran}}
\newacronym{vss}{VSS}{Video Streaming Server}
\newacronym{wbf}{WBF}{Wired Bias Function}
\newacronym{wf}{WF}{Waterfilling}
\newacronym{wg}{WG}{Working Group}
\newacronym{wlan}{WLAN}{Wireless Local Area Network}
\newacronym{osm}{OSM}{Open Source Management and Orchestration}
\newacronym{pnf}{PNF}{Physical Network Function}
\newacronym{drl}{DRL}{Deep Reinforcement Learning}
\newacronym{mtc}{MTC}{Machine-type Communications}
\newacronym{osc}{OSC}{O-RAN Software Community}
\newacronym{mns}{MnS}{Management Services}
\newacronym{ves}{VES}{\gls{vnf} Event Stream}
\newacronym{ei}{EI}{Enrichment Information}
\newacronym{fh}{FH}{Fronthaul}
\newacronym{fft}{FFT}{Fast Fourier Transform}
\newacronym{laa}{LAA}{Licensed-Assisted Access}
\newacronym{plfs}{PLFS}{Physical Layer Frequency Signals}
\newacronym{ptp}{PTP}{Precision Time Protocol}
\newacronym{asic}{ASIC}{Application-specific Integrated Circuit}
\newacronym{aal}{AAL}{Acceleration Abstraction Layer}
\newacronym{fec}{FEC}{Forward Error Correction}
\newacronym{sdl}{SDL}{Shared Data Layer}
\newacronym{nib}{NIB}{Network Information Base}
\newacronym{rnib}{R-NIB}{RAN \gls{nib}}
\newacronym{fcaps}{FCAPS}{Fault, Configuration, Accounting, Performance, Security}
\newacronym{ie}{IE}{Information Element}
\newacronym{fg}{FG}{Focus Group}
\newacronym{osfg}{OSFG}{Open Source Focus Group}
\newacronym{sdfg}{SDFG}{Standard Development Focus Group}
\newacronym{tifg}{TIFG}{Test \& Integration Focus Group}
\newacronym{sfg}{SFG}{Security Focus Group}
\newacronym{swg}{SWG}{Security Work Group}
\newacronym{e2sm}{E2SM}{E2 Service Model}
\newacronym{tsc}{TSC}{Technical Steering Committee}
\newacronym{sdo}{SDO}{Standard-Development Organization}
\newacronym{sql}{SQL}{Structured Query Language}
\newacronym{ssh}{SSH}{Secure Shell}
\newacronym{tls}{TLS}{Transport Layer Security}
\newacronym{netconf}{NETCONF}{Network Configuration Protocol}
\newacronym{dtls}{DTLS}{Datagram Transport Layer Security}
\newacronym{cmp}{CMP}{Certificate Management Protocol}
\newacronym{ccc}{CCC}{Cell Configuration and Control}
\newacronym{dsp}{DSP}{Digital Signal Processing}
\newacronym{opex}{OPEX}{Operational Expenses}
\newacronym{cbrs}{CBRS}{Citizen Broadband Radio Service}
\newacronym{ntn}{NTN}{Non-terrestrial Network}
\newacronym{gbr}{GBR}{Guaranteed Bitrate}
\newacronym{sps}{SPS}{Semi-Persistent Scheduling}
\newacronym{tbs}{TBS}{Transport Block Size}
\newacronym{gnss}{GNSS}{Global Navigation Satellite System}
\newacronym{tof}{ToF}{Time of Flight}
\newacronym{rtof}{RToF}{Return Time of Flight}
\newacronym{rsig}{RS}{Reference Signal}
\newacronym{nrtric}{near-RT RIC}{near-Real Time Ran Intelligent Controller}
\newacronym{nonrtric}{non-RT RIC}{non-Real Time Ran Intelligent Controller}
\newacronym{aoa}{AoA}{Angle of Arrival}
\newacronym{tdoa}{TDoA}{Time Difference of Arrival}
\newacronym{rtoa}{RToA}{Return Time of Arrival}
\newacronym{ecdf}{ECDF}{Empirical Cumulative Distribution Function}
\newacronym{ris}{RIS}{Reconfigurable Intelligent Surface}
\newacronym{srd}{SRD}{Smart Radio Device}
\newacronym{gfbr}{GFBR}{Guaranteed Flow Bit Rate}
\newacronym{rg}{RG}{Resource Grid}
\newacronym{rb}{RB}{Resource Block}
\newacronym{re}{RE}{Resource Element}
\newacronym{rfra}{RF}{Radio Frame}
\newacronym{scs}{SCS}{Subcarrier Spacing}
\newacronym{ec}{EC}{Edge Computing}
\newacronym{af}{AF}{Amplify-and-Forward}
\newacronym{ncr}{NCR}{Network-Controlled Repeater}
\newacronym{tp}{TP}{Test Point}
\newacronym{cs}{CS}{Candidate Site}
\newacronym{src}{SRC}{Smart Radio Connection}
\newacronym{milp}{MILP}{Mixed Integer-Linear Programming}

\newacronym{FCMC}{FCMC}{full coverage minimum cost}

\newacronym{MBCC}{MBCC}{maximum budget-constrained coverage}

\newacronym{PDF}{PDF}{probability density function}
\begin{document}
\title{Advanced Network Planning in  \\6G Smart Radio Environments}

\author{Reza~Agahzadeh~Ayoubi,~\IEEEmembership{Member,~IEEE,}
        Marouan~Mizmizi,~\IEEEmembership{Member,~IEEE,}
        Eugenio~Moro,~\IEEEmembership{Member,~IEEE,}   \\
        Ilario~Filippini,~\IEEEmembership{Senior~Member,~IEEE,}
        Umberto~Spagnolini,~\IEEEmembership{Senior~Member,~IEEE}
        \thanks{This work was partially supported by the European Union - Next Generation EU under the Italian National Recovery and Resilience Plan (NRRP), Mission 4, Component 2, Investment 1.3, CUP D43C22003080001, partnership on “Telecommunications of the Future” (PE00000001 - program “RESTART”)}
\thanks{The authors are with the Department of Electronics, Information and Bioengineering, Politecnico di Milano, 20133, Milano, Italy}}

\maketitle
\pagenumbering{gobble}
\begin{abstract}
The growing demand for high-speed, reliable wireless connectivity in 6G networks necessitates innovative approaches to overcome the limitations of traditional \gls{RAN}. Reconfigurable \gls{RIS} and \gls{NCR} have emerged as promising technologies to address coverage challenges in high-frequency \gls{mmW} bands by enhancing signal reach in environments susceptible to blockage and severe propagation losses. In this paper, we propose an optimized deployment framework aimed at minimizing infrastructure costs while ensuring full area coverage using only \gls{RIS} and \gls{NCR}. We formulate a cost-minimization optimization problem that integrates the deployment and configuration of these devices to achieve seamless coverage, particularly in dense urban scenarios. Simulation results confirm that this framework significantly reduces the network planning costs while guaranteeing full coverage, demonstrating \gls{RIS} and \gls{NCR}’s viability as cost-effective solutions for next-generation network infrastructure.
\end{abstract}

\glsresetall

\section{Introduction}
The demand for data in 6G networks and the advancements in communications require new frequency bands in \gls{mmW} spectra ($24-100$ GHz). Although offering high capacity, these bands have a limited range and are sensitive to blockages, posing challenges to coverage and reliable connectivity. Consequently, next-generation \gls{RAN} designs are evolving to address these limitations with innovative devices such as \gls{RIS}s and \gls{NCR}s, which form the basis of the emerging concept of a \gls{SRE} \cite{SRE}. 

In a \gls{SRE}, the environment becomes an adaptable entity capable of enhancing signal propagation. This is possible by introducing novel network devices, namely \gls{RIS} and \gls{NCR}. \gls{RIS} is a quasi-passive device that manipulates incident electromagnetic (EM) waves, altering their direction through programmable reflection coefficients \cite{GSnell}. Typically structured as two-dimensional arrays of tunable meta-atoms, \gls{RIS}s are deployed to strategically redirect signals and thus improve coverage, especially in urban areas where line-of-sight (LoS) paths are obstructed. In contrast, \gls{NCR} is an active device with beamforming and amplification capabilities and the ability to optimize signal routing \cite{ncr_cell_edge,ncr_deployement}.

Recent studies have assessed the optimal positioning and orientation of \gls{RIS}s and \gls{NCR}s for improved coverage in simplified scenarios. For example, positioning \gls{RIS}s near the transmitter (Tx) or receiver (Rx) has been shown to mitigate path loss, as demonstrated in \cite{RIS_delpoymen_Pos, RIS_delpoymen_Pos2}. Furthermore, studies using stochastic geometry models suggest that \gls{RIS}s enhance coverage, notably in environments with high obstacle density, a common characteristic of urban settings \cite{COV_Analysis_0}. The effect of \gls{RIS} deployment density and interference on performance has also been considered, indicating that increasing the number of \gls{RIS} devices requires careful planning to balance desired and interference levels of signal \cite{bafghi2024stochastic}. The performance of \gls{RIS}s is further influenced by their orientation and phase configuration, which are optimized in studies to improve indoor coverage within shadowed regions \cite{RIS_COV_OPT_INDOOR}.

Recent research on \gls{NCR}s has demonstrated their effectiveness in increasing coverage, especially for cell edge users and managing interference in dense networks \cite{NCR_Interf_Makki}. Comparative analyzes between \gls{RIS}s and \gls{NCR}s suggest that each has unique advantages depending on the specific propagation environment. While large-scale \gls{RIS}s can provide high-capacity links, \gls{NCR}s often perform better when real-time signal amplification is required, as shown in scenarios considering propagation and geometric constraints \cite{RIS_vs_NCR_Reza}.

Network planning for \gls{HSRE} deployment in real environments remains a significant challenge~\cite{Eugenio2}. Previous studies often considered idealized setups, ignoring the constraints imposed by buildings, user mobility, and installation feasibility. For example, planning strategies using simplified urban layouts have shown that the jointly deploying \gls{RIS} and \gls{NCR} can significantly improve reliability and mitigate blockage in urban networks \cite{Eugenio1,RIS_RS_Eugenio,DynamicBlockage}. However, a comprehensive network planning strategy for realistic environments is essential to maximize coverage efficiently and minimize deployment costs.

To address these gaps, this paper presents a practical network planning framework for \gls{HSRE} devices, designed with realistic environmental constraints and advanced planning tools. Specifically, this work combines the physical layer considerations of \cite{RIS_vs_NCR_Reza}, which includes precise channel and propagation modeling, with an optimization approach inspired by the network planning model in \cite{RIS_RS_Eugenio}, though using a more advanced optimization method. Our model minimizes network deployment costs while ensuring full coverage, providing an enhancement over prior planning techniques. It accounts for the physical and geometric characteristics of \gls{RIS}s and \gls{NCR}s, leveraging realistic urban maps to capture the impact of environment-specific factors on network performance. Additionally, we incorporate the statistical effects of static blockages common to urban scenarios and design the model to adapt to dynamic blockages that may emerge in live network operations, thereby supporting robust coverage and connectivity. Notably, we assume a cost model where device costs scale with the configurations of \gls{SRE} components, introducing a practical approach to analyzing deployment costs based on device specifications.

Our findings reveal that optimal planning—specifically the strategic placement and tailored configuration of \gls{RIS} and \gls{NCR} devices—significantly reduces deployment costs while ensuring full coverage. This work demonstrates how careful planning and configuration can meet high connectivity demands in urban environments without excessive infrastructure investment.

The rest of the paper is organized as follows: Section  II introduces the role of \gls{HSRE} in network planning. Section  III details the system model, including \gls{RIS} and \gls{NCR} device descriptions. Section  IV presents the optimization problem, specifying the cost-minimization model for full coverage. Section  V discusses numerical results, deployment insights, and cost analysis. Finally, Section  VI concludes the paper and suggests future research directions.

\section{Smart Radio Environment Model\label{sec:SRE}}

Consider the downlink communication with direct and relayed connections between a \gls{BS} and potential \gls{UE}s, represented by a \gls{tp}. The spatial coordinates for the BS, relay, and \gls{UE} are denoted as $\mathbf{p}_{BS} = [x_{BS}, y_{BS}, z_{BS}]^\mathrm{T}$, $\mathbf{p}_R = [x_R, y_R, z_R]^\mathrm{T}$, and $\mathbf{p}_{UE} = [x_{UE}, y_{UE}, z_{UE}]^\mathrm{T}$, respectively. All potential \gls{UE} locations are collected in $\mathcal{P}=\{\mathbf{p}_{UE}\}$. Both the \gls{BS} and UEs are equipped with antenna arrays, comprising $N_t$ and $N_r$ elements, respectively.
\begin{figure}[t!]
    \centering
    \subfloat[][\small RIS]{\includegraphics[height=0.2\textwidth]{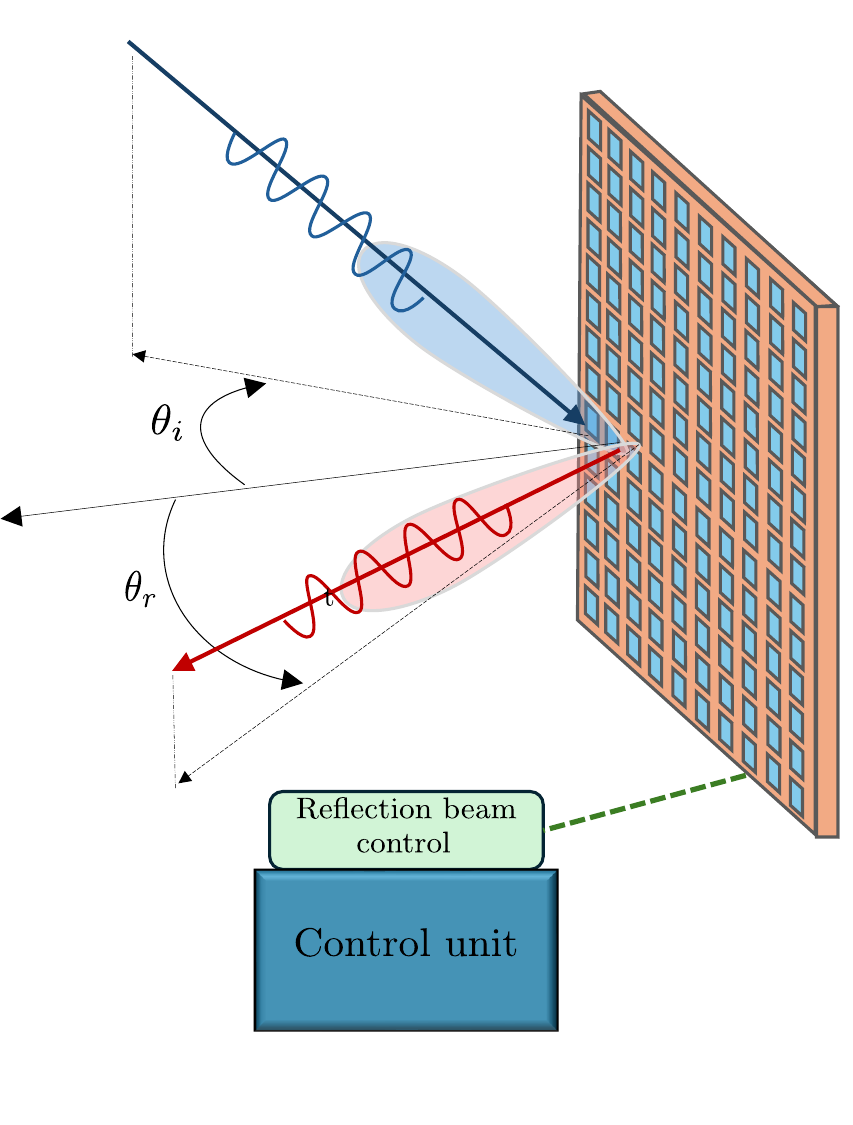}\label{fig:RIS}}\quad
   \
    \subfloat[][\small NCR]{\includegraphics[height=0.2\textwidth]{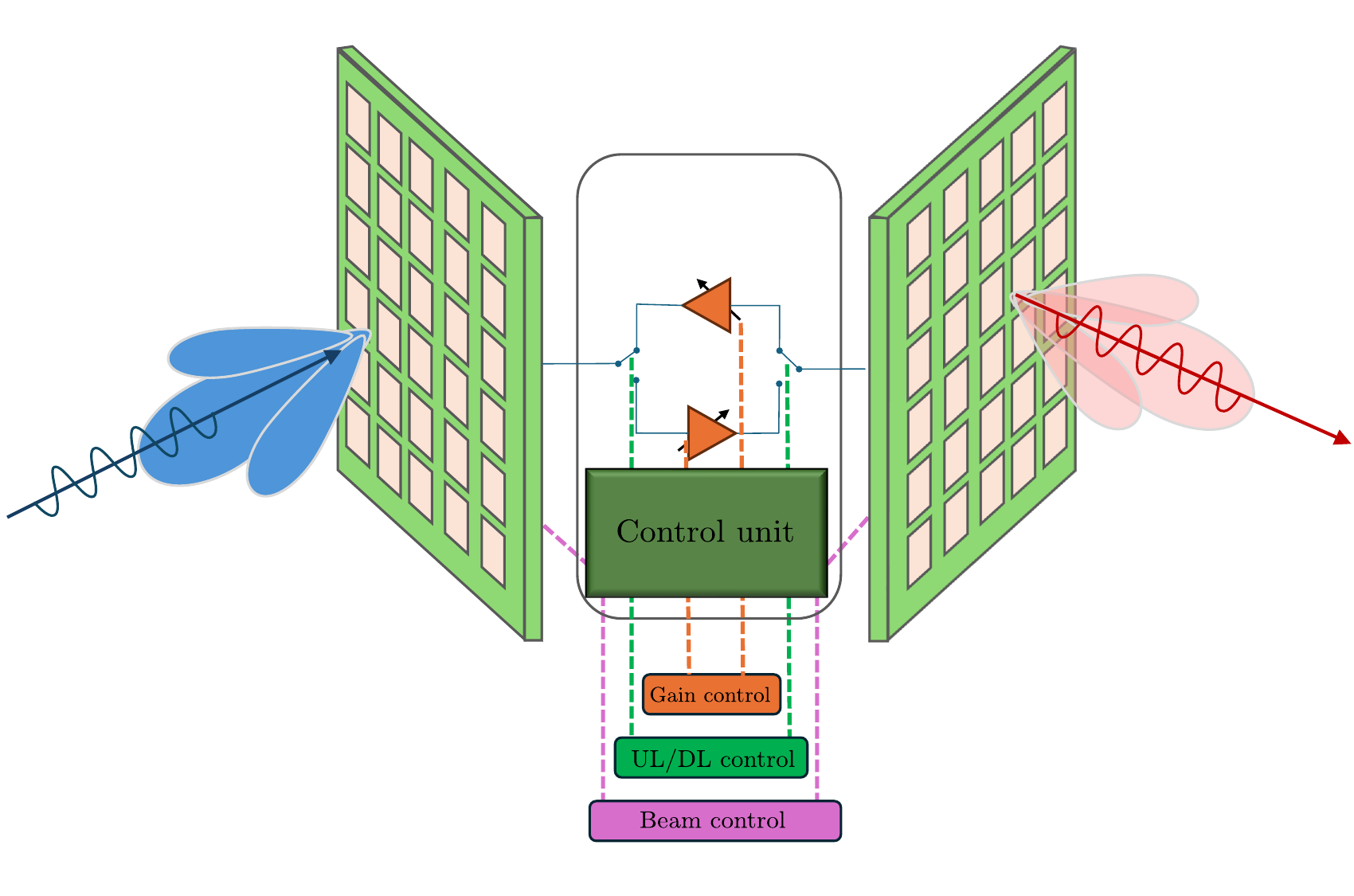}\label{fig:NCR}}
    \caption{Illustrative scheme of different \gls{SRE} components}
\end{figure}
The relay devices considered in this work include the \gls{RIS} and the \gls{NCR}.

The \gls{RIS} is constructed as a planar metasurface comprising \( M \) sub-wavelength meta-atoms, each controlled by a PIN diode or varactor for phase manipulation, as shown in Figure \ref{fig:RIS}. This configuration enables dynamic reconfiguration and directional reflection, enhancing coverage in obstructed areas. The cost model for an \gls{RIS} is structured with both a fixed deployment cost and a variable cost that scales linearly with the number of meta-atoms. This linear scaling reflects the increased number of control components (diodes or varactors) and additional circuitry needed as RIS size grows \cite{SRE}, making the cost proportional to the number of meta-atoms. We model the total cost of an \gls{RIS} as:
\begin{equation}
O^{\mathrm{RIS}} = O^{\mathrm{RIS}}_d + O^{\mathrm{RIS}}_c \cdot M,
\label{eq:ris_cost}
\end{equation}
where \( O^{\mathrm{RIS}}_d \) is the initial deployment cost, and \( O^{\mathrm{RIS}}_c \) represents the unit cost per meta-atom.

The \gls{NCR} configuration, depicted in Figure \ref{fig:NCR}, consists of two antenna panels, each with \( N_p \) elements. The first panel is oriented toward the \gls{BS}, while the second panel, facing the target coverage area, is positioned at an angular separation \( \alpha \) to prevent loop-back interference, which is essential for effective signal relaying through beamforming and amplification. Operating in TDD mode, \gls{NCR}s extend signal reach and manage interference, with continuous power consumption that adds to the overall operating cost. Unlike the passive \gls{RIS}, \gls{NCR}s require ongoing power for signal amplification and active beamforming, which increases the operational expenditure (OPEX) beyond the initial deployment costs (CAPEX) \cite{SRE}. Given these factors, the total cost of an \gls{NCR} includes both a deployment cost and a variable power consumption cost, modeled as a function of its amplification gain \( g \):
\begin{equation}
O^{\mathrm{NCR}} = O^{\mathrm{NCR}}_d + O^{\mathrm{NCR}}_g \cdot [g]_{dB},
\label{eq:ncr_cost}
\end{equation}
where \( O^{\mathrm{NCR}}_d \) denotes the deployment cost, and \( O^{\mathrm{NCR}}_g \) represents the cost per dB of gain. The cost models, will be used in this paper for networks planning optimization, as in \cite{Eugenio2}.

To clarify, the cost models used here are rational estimates based on each device's component and operational needs. Prices are represented in relative units, with a 100 × 100 \gls{RIS} taken as the baseline at 1 unit. Other device prices are scaled accordingly, emphasizing that relative rather than absolute costs drive the optimization. While actual costs vary with market conditions and production scales, our framework can seamlessly adapt to precise vendor data as it becomes available. Any adjustments in the cost model may affect specific results but do not alter the optimization approach itself, which is independent of exact pricing assumptions.

Antenna arrays are arranged with half-wavelength spacing at the carrier frequency $f_0=c/\lambda$, where $c$ is the speed of light, and $\lambda$ is the wavelength. The meta-atoms on the \gls{RIS} are spaced at intervals of $\lambda/4$.

\subsection{Signal Model}\label{subsect:signal_model}

Let $s \in \mathbb{C}$ be the complex symbol to be transmitted, where $\mathbb{E}[s^*s] = \sigma_s^2$, representing the transmitted power. The transmitted signal is then expressed as
\begin{equation}\label{eq:txSignal}
    \mathbf{x} = \mathbf{f}\,s,
\end{equation}
where $\mathbf{f}$ denotes the precoding vector, such that $\|\mathbf{f}\|_\mathrm{F}^2 = N_t$. The transmitted signal in \eqref{eq:txSignal} is received by the \gls{UE} directly $\mathbf{r}^\mathrm{d}$, through an \gls{NCR} $\mathbf{r}^\mathrm{NCR}$, or via a \gls{RIS} $\mathbf{r}^\mathrm{RIS}$. This is represented as:
\begin{equation}\label{eq:rxTotal}
    y = \mathbf{w}^\mathrm{H} \left(\mathbf{r}^\mathrm{d} + \sum_{i\in \mathcal{I}} \mathbf{r}_i^\mathrm{RIS} + \sum_{j\in \mathcal{J}} \mathbf{r}_j^\mathrm{NCR}\right) + \mathbf{w}^\mathrm{H} \, \mathbf{n},
\end{equation}
where $\mathbf{n} \sim \mathcal{CN}\left(\mathbf{0}, \sigma_n^2 \, \mathbf{I}_{N_r}\right)$ represents additive white Gaussian noise, $\mathbf{w} \in \mathbb{C}^{N_r \times 1}$ is the combining vector, and $\mathcal{I}, \mathcal{J}$ denote the sets of deployed \gls{RIS}s and \gls{NCR}s, respectively.

\subsubsection{Direct Signal Model}
The received signal for the direct link is:
\begin{equation}\label{eq:rxDirect}
    \mathbf{r}^\mathrm{d}  =  \mathbf{H}^\mathrm{d} \, \mathbf{x}
\end{equation}
with $\mathbf{H}^\mathrm{d} \in \mathbb{C}^{N_r \times N_t}$ representing the MIMO direct channel.

\subsubsection{Relayed Signal Model via \gls{RIS}}
The received signal for the relayed link through an \gls{RIS} is expressed as:
\begin{equation}\label{eq:rxRIS}
    \mathbf{r}^\mathrm{RIS} = \mathbf{H}_o^\mathrm{RIS} \, \boldsymbol{\Phi} \, \mathbf{H}_i^\mathrm{RIS} \, \mathbf{x},
\end{equation}
 with $\boldsymbol{\Phi} \in \mathbb{C}^{M\times M}$ as the reflection coefficient matrix. Here, $\mathbf{H}_i^\mathrm{RIS} \in \mathbb{C}^{M\times N_t}$ and $\mathbf{H}_o^\mathrm{RIS} \in \mathbb{C}^{N_r\times M}$ are the input/output channel matrices between the BS, the \gls{RIS}, and the UE. The reflection matrix $\boldsymbol{\Phi}$ in \eqref{eq:rxRIS} is diagonal with entries defined as:
\begin{equation}\label{eq:reflectingmatrix}
    \boldsymbol{\Phi} = \text{diag}\left( e^{j\phi_{1}},...,e^{j\phi_{m}},..., e^{j\phi_{M}}\right),
\end{equation}
where $\phi_{m}$ is the phase shift applied at the $m$-th element.

\subsubsection{Relayed Signal Model via \gls{NCR}}
Since the \gls{NCR} is an active device, the received signal model includes amplified noise. The signal received by the \gls{NCR} is:
\begin{equation}\label{eq:rxNCR}
    \mathbf{z}^\mathrm{NCR} = \mathbf{H}_i^\mathrm{NCR} \, \mathbf{x} + \mathbf{v},
\end{equation}
where $\mathbf{v} \in \mathbb{C}^{N_p \times 1} \sim \mathcal{CN}\left(\mathbf{0}, \sigma_v^2 \, \mathbf{I}_{N_p}\right)$ represents the noise at the \gls{NCR}, and $\mathbf{H}_i^\mathrm{NCR} \in \mathbb{C}^{N_p \times N_t}$ is the channel matrix between the \gls{BS} and the \gls{NCR}. The signal $\mathbf{z}^\mathrm{NCR}$ in \eqref{eq:rxNCR} is amplified and forwarded towards the UE. The signal received by the \gls{UE} via the \gls{NCR} can be expressed as:
\begin{equation}
    \mathbf{r}^\mathrm{NCR} = \sqrt{g} \, \mathbf{H}_o^\mathrm{NCR} \, \mathbf{Q} \, \mathbf{z}^\mathrm{NCR},
\end{equation}
where $g$ denotes the amplification gain, $\mathbf{H}_o^\mathrm{NCR} \in \mathbb{C}^{N_r \times N_p}$ is the channel matrix between the \gls{NCR} and the UE. The relaying matrix $\mathbf{Q} \in \mathbb{C}^{N_p \times N_p}$ applies a phase shift between the received and forwarded signals. For the \gls{NCR} configuration, the relaying matrix is:
\begin{equation}
    \mathbf{Q} = \mathbf{u} \, \mathbf{b}^\mathrm{H},
\end{equation}
where $\mathbf{u} \in \mathbb{C}^{N_p \times 1}$ and $\mathbf{b} \in \mathbb{C}^{N_p \times 1}$ represent the forwarding and receiving beamformers, respectively.

\subsection{Channel Model}\label{subsect:channel_model in presence of RIS}

We assume a block-fading channel model with independent fading among direct $\mathbf{H}^\mathrm{d}$, forward $\mathbf{H}_i$, and backward $\mathbf{H}_o$ channels for \gls{NCR} and \gls{RIS}. Given the challenging propagation conditions at mmWave frequencies, we adopt the Saleh-Valenzuela cluster-based model (see \cite{Molisch}). The impulse response of any channel in the model can be expressed as:
\begin{equation}\label{eq:channelModel}
    \mathbf{H} = \sum_{p=1}^{P} \frac{\alpha_p}{\sqrt{P}}\, \varrho_r(\vartheta_p^r, \varphi_p^r) \varrho_t(\vartheta_p^t, \varphi_p^t) \mathbf{a}_r(\vartheta_p^r, \varphi_p^r)\mathbf{a}_t(\vartheta_p^t, \varphi_p^t)^\mathrm{H},
\end{equation}
where $P$ is the number of paths, $\alpha_p$ denotes the scattering amplitude of the $p$-th path, and $\mathbf{a}_t(\vartheta_p^t, \varphi_p^t)\in\mathbb{C}^{N_t\times 1}$ and $\mathbf{a}_r(\vartheta_p^r, \varphi_p^r)\in\mathbb{C}^{N_r\times 1}$ denote the Tx and Rx array response vectors for the $p$-th path, which depend on the Tx and Rx pointing angles $(\vartheta_p^t, \varphi_p^t)$ and $(\vartheta_p^r, \varphi_p^r)$. Element radiation patterns $\varrho(\vartheta_p, \varphi_p)$ are incorporated according to \cite{3GPP} for BS, UE, and \gls{NCR}, and are defined as in \cite{CIRS} for the \gls{RIS}.

Assuming a uniform planar array, the array factor $\mathbf{a}(\vartheta, \varphi)$ in \eqref{eq:channelModel} can be expressed as:
\begin{equation}
    \mathbf{a}(\vartheta, \varphi) = [e^{j\mathbf{k}(\vartheta, \varphi)^\mathrm{T} \boldsymbol{\nu}_1}, \cdots, e^{j\mathbf{k}(\vartheta, \varphi)^\mathrm{T} \boldsymbol{\nu}_L}],
\end{equation}
where $\mathbf{k}(\vartheta, \varphi) \in \mathbb{R}^{3 \times 1}$ is the wave vector, defined as:
\begin{equation}
    \mathbf{k}(\vartheta, \varphi) = \frac{2\pi}{\lambda}[\cos(\varphi)\cos(\vartheta), \cos(\varphi)\sin(\vartheta), \sin(\varphi)]^T,
\end{equation}
and $\boldsymbol{\nu}_\ell = [x_\ell, y_\ell, z_\ell]$ represents the position of the $\ell$-th element in the array, in local coordinates.

Using the calculated direct or relay channels, we can compute the instantaneous \gls{SNR} for each link, denoted by $\gamma_0$, following the approach in \cite{RIS_vs_NCR_Reza}. Static blockage is handled deterministically using actual building maps to form the scenario, while the dynamic blockage model from \cite{DynamicBlockage} is used for long-term \gls{SNR} calculations. For example, for the direct path, the long-term \gls{SNR} is calculated as in \cite{RIS_vs_NCR_Reza}:
\begin{align}
    &\overline{\gamma}^\textrm{BS} = \mathrm{P}^{\textrm{BS}}_B\,\gamma^{\textrm{BS}} + (1-\mathrm{P}^{\textrm{BS}}_B)\gamma_{0}^\textrm{BS},\label{eq:LT_SNR_direct}
\end{align}
where $\mathrm{P}^{\textrm{BS}}_B$ is the blockage probability of the direct paths, $\gamma_{0}^\textrm{BS}$ is the \gls{SNR} upper limit when blockage does not occur, and ${\gamma}^\textrm{BS}$ is the \gls{SNR} accounting for penetration loss or knife-edge diffraction \cite{RIS_vs_NCR_Reza, 3GPPTR38901}. The long-term \gls{SNR} for relayed links is calculated similarly. The dynamic blockage probability $\mathrm{P}_B$ depends on link length, as well as parameters like blocker density, velocity, and dimensions \cite{DynamicBlockage}. These long-term \gls{SNR}s are used by the optimization models presented in the next section.

\section{Cost Minimization}

The \gls{mmW} \gls{SRE}-based networks depend significantly on deployment geometry, which directly impacts performance, as shown in previous planning studies~\cite{RIS_RS_Eugenio}. Due to the potential for large-scale deployments resulting from the cost-efficiency of these devices, understanding system-level impacts is essential. To precisely assess network performance with these devices integrated, we employ a \gls{milp} optimization approach. This approach ensures an optimal deployment layout, providing performance insights that reflect the best possible outcomes for networks using \gls{RIS} and \gls{NCR} devices. The objective is to minimize network costs while ensuring full coverage.

\subsection{System Model}
\newcommand{\R}{\mathcal{R}}
\newcommand{\N}{\mathcal{N}}

Consider a geographic area where the coverage of a pre-deployed \gls{BS} requires enhancement through \gls{SRE} devices. Let $\mathcal{C}$ represent the set of \glspl{cs} for potential \gls{SRE} device installations.

This work considers two types of \gls{SRE} devices, \gls{RIS} and \gls{NCR}, with distinct performance metrics and costs. The \gls{milp} approach optimizes network planning by choosing between these device types at each CS to maximize planning objectives. This is mathematically modeled with the set $\mathcal{D}$, which represents available technologies (\gls{RIS} or \gls{NCR}) and configurations, such as \gls{RIS} size or \gls{NCR} amplification gain.

Let $\mathcal{T}$ denote the set of \glspl{tp}, representing sampled positions in the geographic area where coverage is evaluated. Each \gls{tp} $t \in \mathcal{T}$ is considered covered if the \gls{SNR} at that location exceeds a threshold $\Gamma$. This threshold, representing the minimum guaranteed \gls{SNR} enforced at each \gls{tp}, is determined prior to planning optimization and shapes the final network topology.
\begin{figure*}[thb!]
\centering
\subfloat[][\small $K=1$, $\Gamma = 0$ dB ]
{\includegraphics[width=0.32\textwidth]{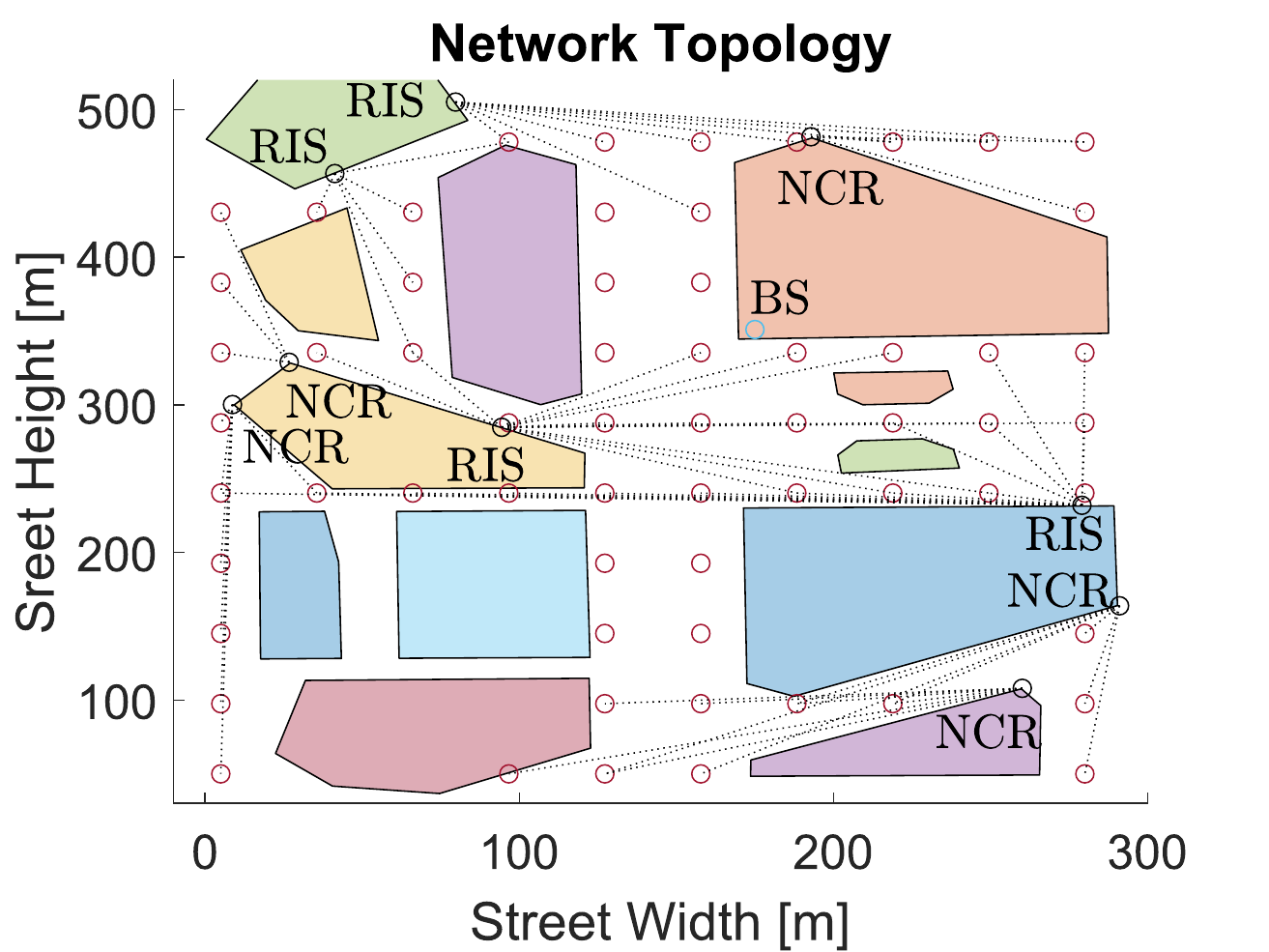}}
\subfloat[][\small  $K=1$, $\Gamma = 20$ dB]
{\includegraphics[width=0.32\textwidth]{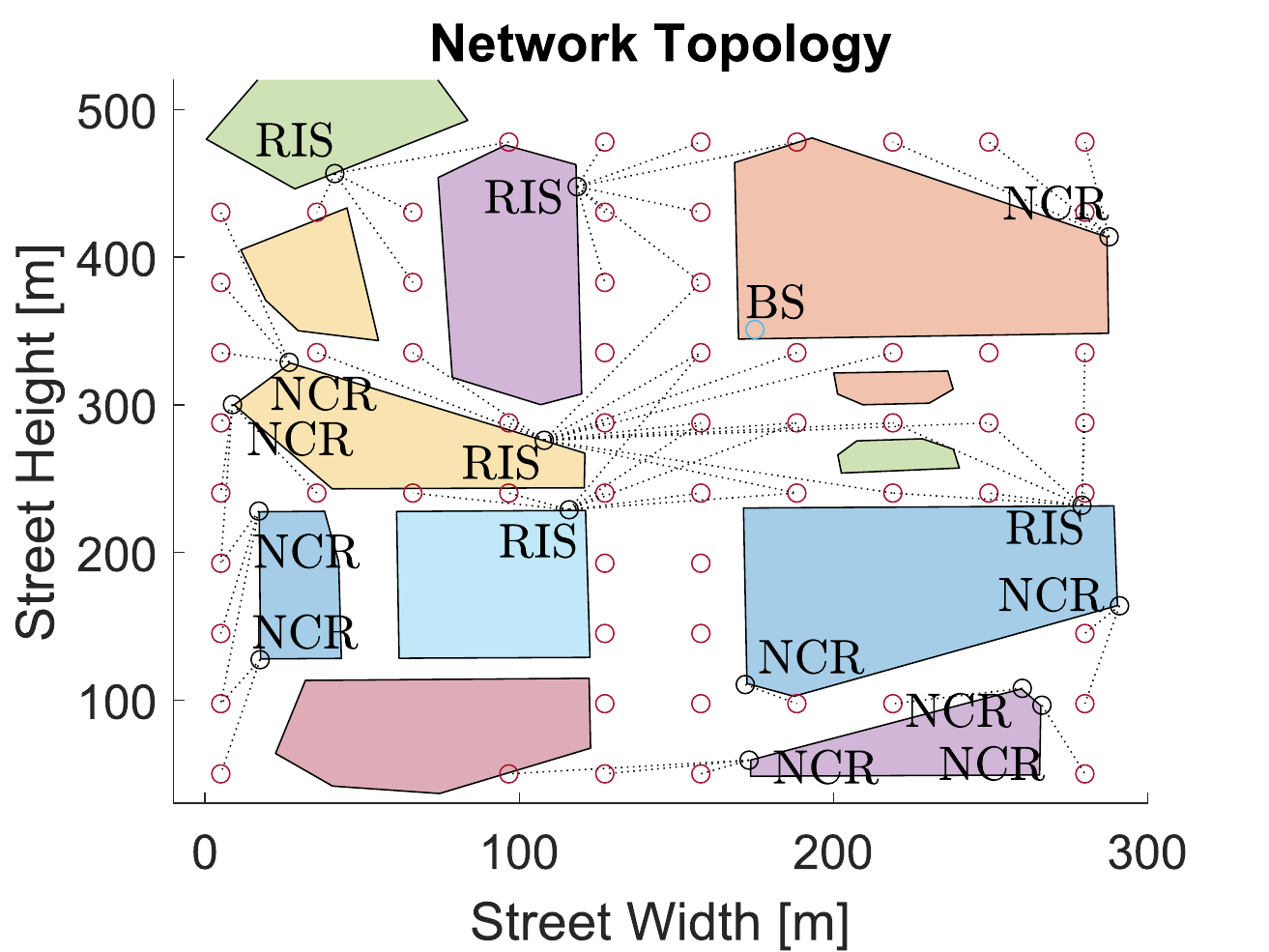}}
\subfloat[][\small  $K=2$, $\Gamma = 0$ dB]
{\includegraphics[width=0.32\textwidth]{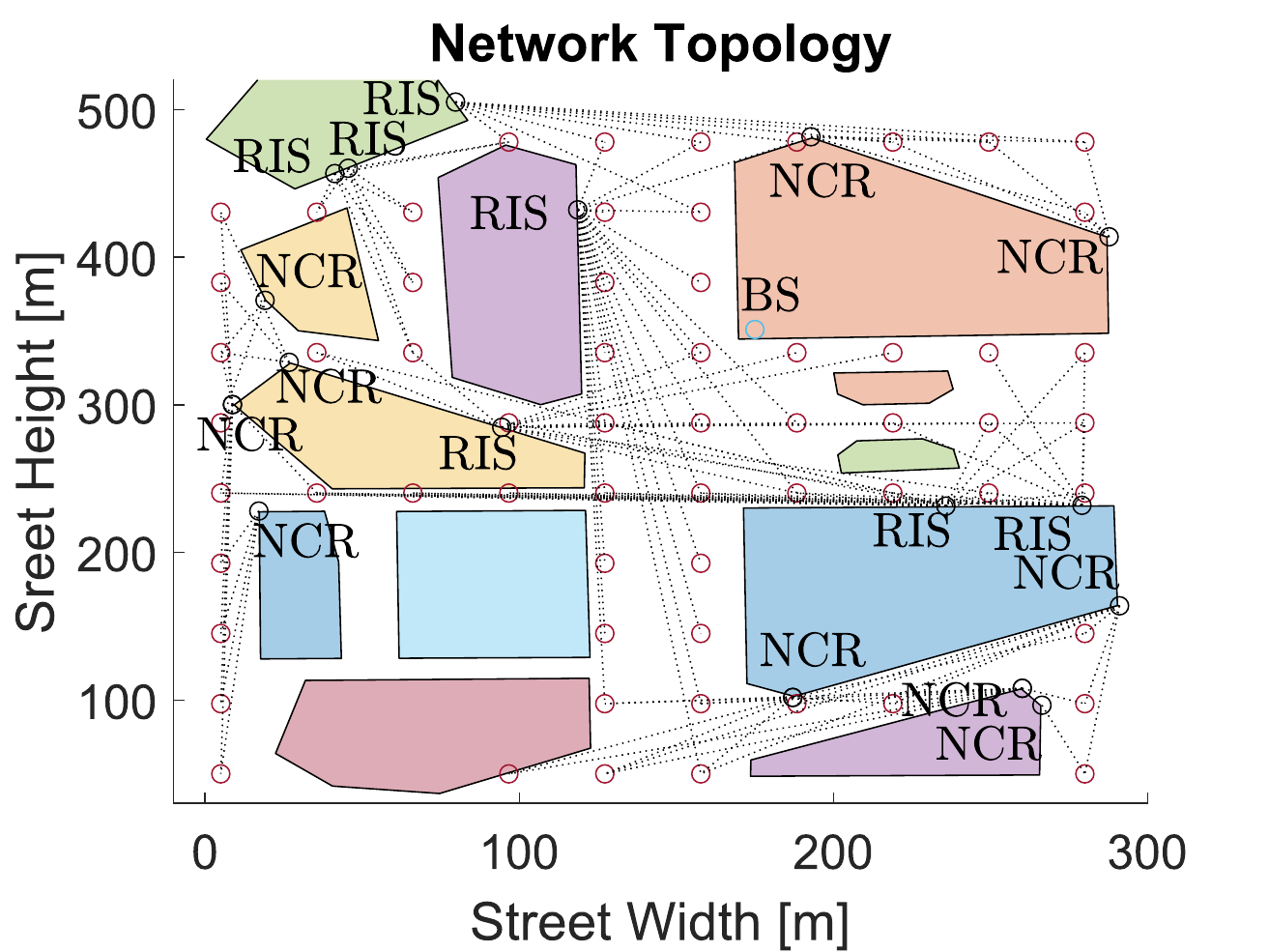}}
    \caption{ Optimally planned network topologies for minimizing total installation cost, for the scenario of Piazza Piola in Milan \ref{tab:SimParam_1}, varying the  the guaranteed number of connections for each test point $K$ and \gls{SNR} threshold in dB $\Gamma$, using the default parameters in Table \ref{tab:SimParam_1}.}\label{fig:map_planning}
\end{figure*}

The measured \gls{SNR} depends on the final network topology and active \gls{SRE} devices. Each \gls{tp} $t \in \mathcal{T}$ can be served either directly by the \gls{BS} or indirectly through an \gls{SRE} device. Let $\overline{\gamma}^\text{BS}_t$ represent the long-term \gls{SNR} measured at \gls{tp} $t$ from the \gls{BS} only. Using the channel model from Sect.~\ref{subsect:channel_model in presence of RIS}, we define a boolean activation parameter $\Delta_{t}^\text{BS}$ as follows:
\begin{equation}
  \Delta_t^\text{BS}=\left\{
  \begin{array}{@{}ll@{}}
    1, & \text{if}\ \overline{\gamma}_t^\text{BS} \geq \Gamma \\
    0, & \text{otherwise.}
  \end{array}\right.
  \label{eq:bs_act_param}
\end{equation}
This indicates that $\Delta_{t}^\text{BS}$ is 1 if the \gls{BS} can meet the \gls{SNR} threshold at \gls{tp} $t$, and 0 otherwise.

Similarly, we calculate the \gls{SNR} at each \gls{tp} from active \gls{SRE} devices, defining $\overline{\gamma}_{t,c}^{d}$ as the \gls{SNR} at \gls{tp} $t$ from a device of type $d \in \mathcal{D}$ at CS $c \in \mathcal{C}$. The activation parameter $\Delta_{t,c}^{d}$ is defined as:
\begin{equation}
  \Delta_{t,c}^d=\left\{
  \begin{array}{@{}ll@{}}
    1, & \text{if}\ \overline{\gamma}_{t,c}^{d} \geq \Gamma \\
    0, & \text{otherwise}
  \end{array}\right.
  \label{eq:ris_act_param}
\end{equation}

These activation parameters ($\Delta_t^\text{BS}$ and $\Delta_{t,c}^d$) inform the optimization model about feasible coverage links. They can also be adjusted to account for obstructions or other deployment constraints.

\subsection{Optimization Formulation}

During the optimization phase, we define $O^{d}$ as the cost of installing a device of type $d \in \mathcal{D}$. We then define the decision variable $v_{c}^{d} \in \{0,1\}$, where $v_{c}^{d}=1$ if a device of type $d$ is installed at CS $c$, and $v_{c}^{d}=0$ otherwise. The optimization formulation is as follows:

\begin{subequations}
\begin{align}
     \quad &\min \sum_{c \in \mathcal{C}, d \in \mathcal{D}} O^d v_c^d \label{opt:fcmc:obj} \\
    \text{s.t.} \quad 
    &\Delta_{t}^\text{BS} + \sum_{c \in \mathcal{C}, d \in \mathcal{D}} \Delta_{t,c}^d v_{c}^d \geq K 
    \quad \forall t \in \mathcal{T} \label{opt:fcmc:cov} \\
    &v_c^d \in \{0,1\} \quad \forall c \in \mathcal{C}, d \in \mathcal{D} \label{opt:fcmc:binary}
\end{align}\label{opt:fcmc}
\end{subequations}

In~(\ref{opt:fcmc:obj}), the objective function minimizes network costs by optimizing device installations. Constraint~(\ref{opt:fcmc:cov}) ensures each \gls{tp} is covered by at least $K$ distinct devices, achieving the required \gls{SNR}. Although computationally intensive, this formulation effectively addresses realistic planning scenarios, as discussed in Sect.~\ref{sec:results}.

\section{Results}
\label{sec:results}
\begin{table}[b!]
    \centering
    \footnotesize
    \caption{Default simulation parameters used in Section  \ref{sec:results}.}
    \begin{tabular}{l|c|c}
    \toprule
        \textbf{Parameter} &  \textbf{Symbol} & \textbf{Value(s)}\\
        \hline
        Carrier frequency & $f_0$  & $28$ GHz \\
        Bandwidth & $B$ & $200$ MHz\\
        \gls{BS} Transmit power & $\sigma^2_s$ & $35$ dBm\\
        Noise power & $\sigma^2_n$ & $-82$ dBm\\
        \gls{NCR} array size & $N_{p,h}\times N_{p,v}$ & $12 \times 6$ \\
        \gls{NCR} amplification gain & $\vert  g\vert^2$           & $55$ dB\\ 
        \gls{RIS} elements & $\sqrt{M} \times \sqrt{M}$ & $100 \times 100$\\
        \gls{RIS} element spacing & $d_n,d_m$ & $\lambda_0/4$ m\\
        \gls{BS} antenna array & $N_\mathrm{t} $ & $12\times 16$ \\
        Device prices & $O^\mathrm{ris}$, $O^\mathrm{ncr}$ & 1, 3  \\
Blocker height  \cite{DynamicBlockage}                                                    & $h_{B}$         & 1.7 m
\\ 
Blocker density \cite{DynamicBlockage}                                                     & $\lambda_{B}$           & $4\times 10^{-3}$ m$^{-2}$\\
Blocker velocity \cite{DynamicBlockage}                           & $V$           & 15 m/s
\\
Blockage duration \cite{DynamicBlockage}                           & $\mu^{-1}$            & 5 s
\\ 
        \bottomrule
    \end{tabular}
    \label{tab:SimParam_1}
\end{table}
In this section we provide numerical results, achieved with the optimization model \eqref{opt:fcmc}. For simplicity, we assume a uniform building height of 6 meters due to limited data on specific heights. Building shapes are convexified as in \cite{U6G}, preserving spatial accuracy while streamlining the network planning process. \gls{RIS} \glspl{cs} are positioned at a height of 5 meters on building walls, while \gls{NCR} \glspl{cs} are set at 6.5 meters, reflecting the rooftop placement. The candidate sites (\glspl{cs}) for \gls{RIS}s are spaced at regular intervals of 5 meters along building walls. For \gls{NCR}s, \glspl{cs} are chosen from rooftop vertices, with one panel facing the \gls{BS} and the other directed toward target coverage areas. The default simulation parameters are provided in Table \ref{tab:SimParam_1}, unless specified otherwise. 

We begin by presenting the optimally planned network topologies for an exemplary scenario at Piazza Piola in Milan, illustrated in Figure \ref{fig:map_planning}. To examine the impact of varying network requirements, we adjust the guaranteed number of connections for each \gls{tp}, denoted as $K$, and the \gls{SNR} threshold, $\Gamma$, measured in dB. The specific values of these parameters are detailed in the figure captions.

In the scenario with $K=1$ and $\Gamma=0$ dB in Figure (a), a total of 4 \gls{RIS}s and 5 \gls{NCR}s are deployed, resulting in a total cost of 19 units. Increasing the \gls{SNR} threshold to 20 dB in Figure (b) prompts the optimization model to deploy 5 \gls{RIS}s and 10 \gls{NCR}s, raising the total cost to 35 units. Alternatively, increasing the number of paths each \gls{tp} can be served by to $K=2$ (Figure (c)) results in the deployment of 7 \gls{RIS}s and 10 \gls{NCR}s, with a total cost of 37 units.

In each scenario, not only the number of devices changes, but also most of the times, the model chooses different \gls{cs}s to install them.  These results indicate that as the \gls{SNR} threshold increases, the model tends to prioritize \gls{NCR} installations due to their amplification gain, which better supports higher \gls{SNR} demands.

In the next step, to obtain broader, non-site-specific results, we apply the optimization model across eight distinct $400 \times 400$ m regions within the Milan map. In each region, a single \gls{BS} is positioned near the center on top of a building.

\begin{figure}[!htb]
\centering
{\includegraphics[width=0.45\textwidth]{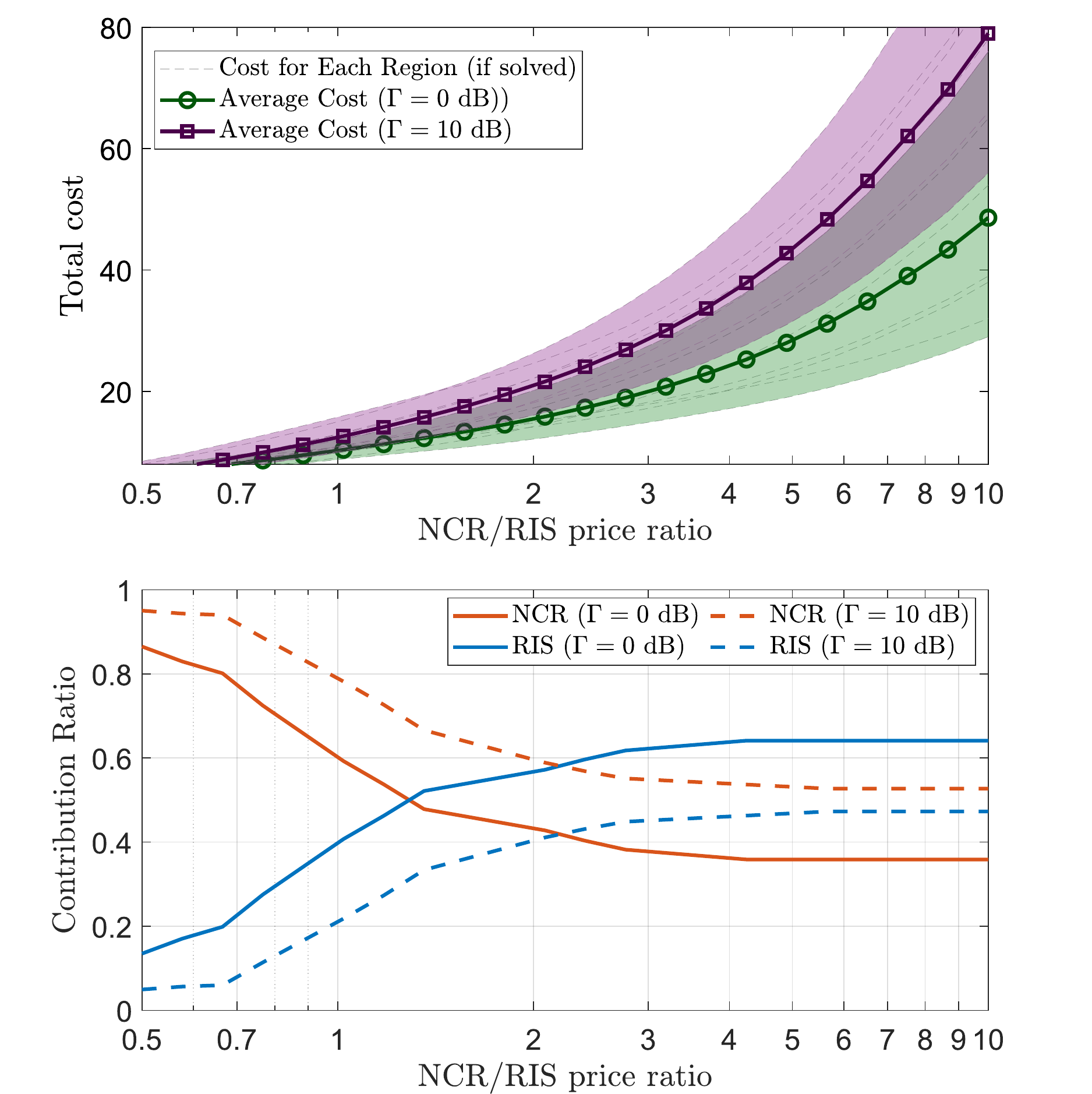}}
 \caption{Average total cost vs \gls{NCR}/\gls{RIS} price ratio with FCMC optimization model, for fixed settings of \gls{RIS}/\gls{NCR} (see Table \ref{tab:SimParam_1}).}\label{fig:opt2_ratio}
\end{figure}

To begin, we fix the default configurations for both devices and examine how the total cost of achieving 100\% coverage of the \glspl{tp}s scales with the price ratio of \gls{NCR} to \gls{RIS}, represented as $\ O ^{\mathrm{NCR}} / \ O ^{\mathrm{RIS}}$. Figure \ref{fig:opt2_ratio} illustrates the total cost versus the \gls{NCR}-to-\gls{RIS} price ratio, with device configurations set to the default values in Table \ref{tab:SimParam_1}. The gray dashed lines in the figures denote each region, while shaded areas represent the upper and lower bounds derived from these regions. The thick solid lines with markers indicate the average values across scenarios.

As the price ratio increases, the total cost for each region, along with the average total cost, shows an exponential trend. This trend is particularly pronounced due to the requirement that all \glspl{tp} be served. For \glspl{tp} located at a distance from the \gls{BS} or obstructed by buildings, service can only be provided by \gls{NCR} devices, which, despite their higher cost, are essential for extending coverage in such challenging areas. Specifically, with \gls{SNR} threshold $\Gamma = 0$ dB, if the cost ratio exceeds $\ O ^{ncr}/\ O ^{ris}>1.3$, the contribution of \gls{RIS}s surpasses that of \gls{NCR}s. For \gls{SNR} threshold $\Gamma = 10$ dB, even when the ratio exceeds $\ O ^{ncr}/\ O ^{ris}>3$, the contribution of \gls{NCR}s remains higher than that of \gls{RIS}s, due to the high \gls{SNR} requirements that only \gls{NCR}s can satisfy, regardless of price.

Having fixed the default configurations of the devices in the previous example, in the next examples, we scale the price of \gls{RIS}, according to the relation \eqref{eq:ris_cost} with the initial deployment cost $O^{\mathrm{RIS}}_d=0.4$, and cost per meta-atom $O^{\mathrm{RIS}}_c=6\times10{-5}$, and we scale the price of \gls{NCR} according to relation \eqref{eq:ncr_cost} with the deployment cost $O^{\mathrm{NCR}}_d=0.8$ , and cost per dB of gain $O^{\mathrm{NCR}}_g= 4\times 10^{-2}$. As mentioned in Section \ref{sec:SRE}, these values are assumed such that a \gls{RIS} with a default dimension of $M=100\times 100$ costs 1 unit, while a \gls{NCR} with default parameters would cost more expensive. The rational behind such models lies in their used components \cite{SRE}.

\begin{figure}[!htb]
\centering
{\includegraphics[width=0.45\textwidth]{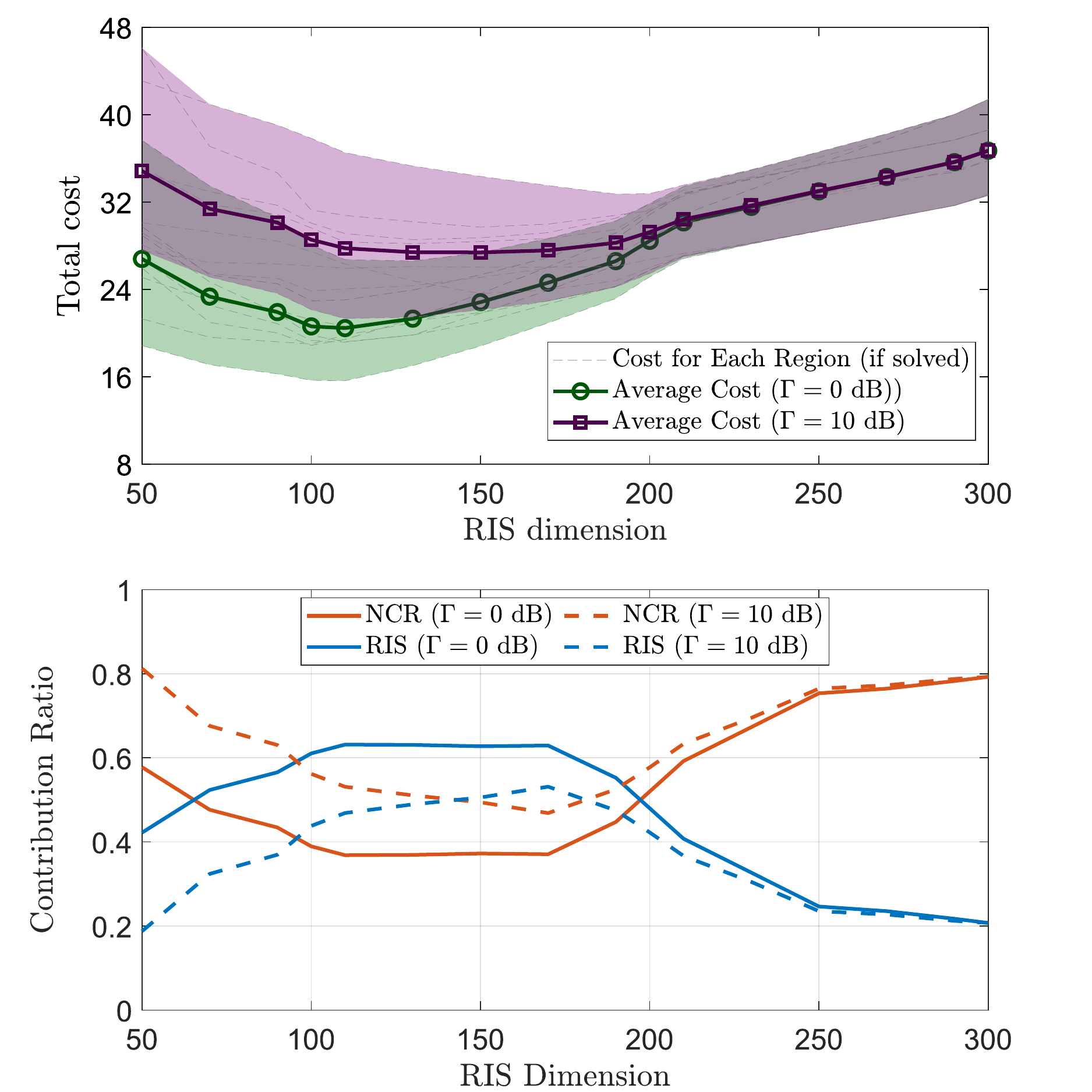}}
 \caption{Average total cost vs \gls{RIS} dimension where the configurations and price of the \gls{NCR} is the default according to Table \ref{tab:SimParam_1} and the cost of the \gls{RIS} scales according to relation \eqref{eq:ris_cost}.}\label{fig:opt2_RIS_dim}
\end{figure}

We first vary the \gls{RIS} configuration, keeping the \gls{NCR} gain fixed. Figure \ref{fig:opt2_RIS_dim} shows the total cost versus \gls{RIS} dimension, $\sqrt{M}$. For RIS configurations ranging from $M = 50 \times 50$ to $M = 150 \times 150$, full coverage is achievable with reduced cost, although increasing dimensions beyond certain points begins to raise costs. Optimized configurations show approximately 20\% cost reduction in certain cases.

\begin{figure}[!htb]
\centering
{\includegraphics[width=0.45\textwidth]{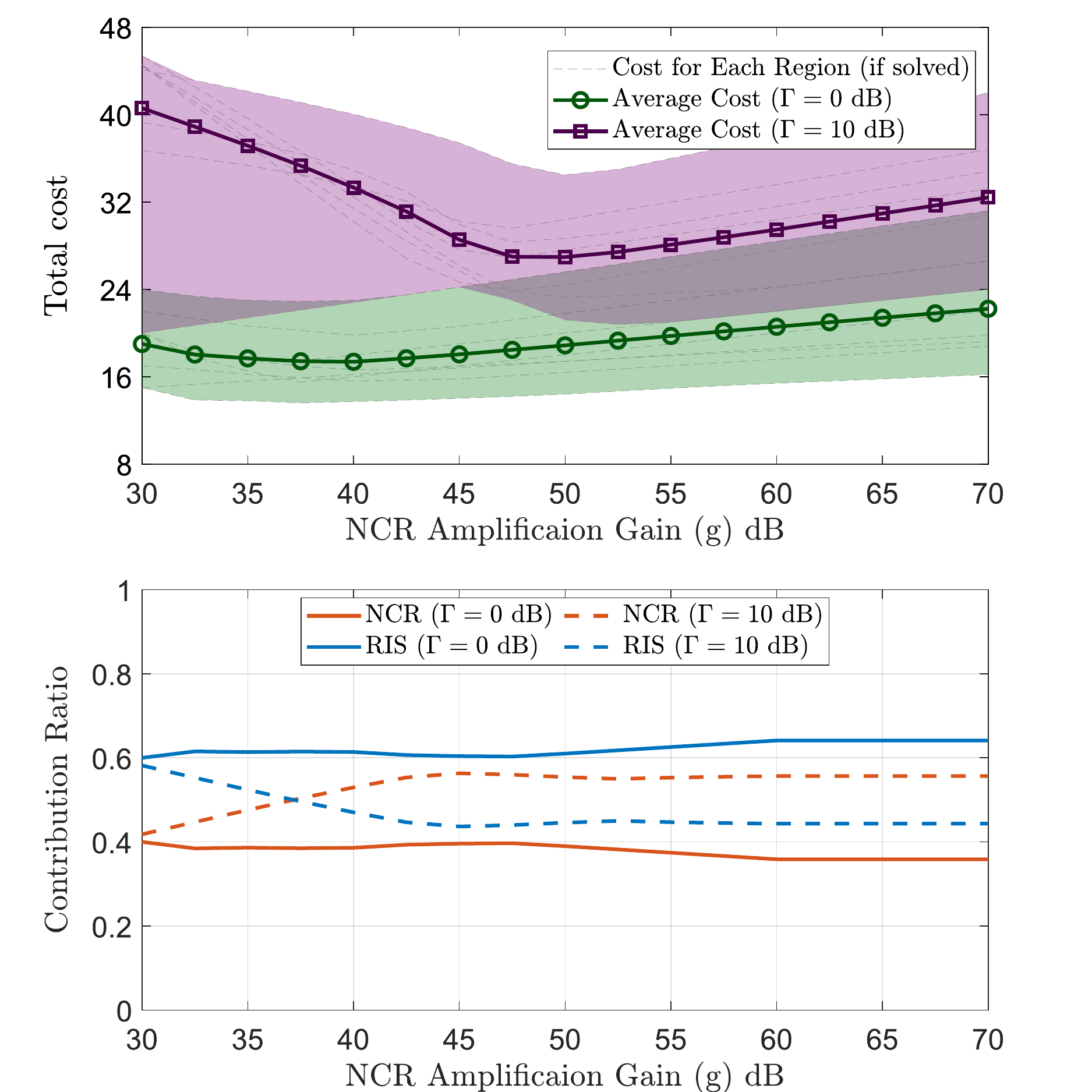}}
\caption{Average total cost vs \gls{NCR} gain, where the configuration and price of the \gls{RIS} is the default according to Table \ref{tab:SimParam_1} and the cost of the \gls{NCR} scales according to relation \eqref{eq:ncr_cost}.}\label{fig:opt2_ncr_gain}
\end{figure}

Similarly, we examine how \gls{NCR} gain impacts total cost, keeping \gls{RIS} configuration fixed. Figure \ref{fig:opt2_ncr_gain} shows that the minimum cost occurs around $g = 38$ dB for $\Gamma = 0$ dB, and $g = 48$ dB for $\Gamma = 10$ dB. Gains above optimal levels only increase total costs without improving coverage. This trend demonstrates that efficient cost management can be achieved by balancing \gls{RIS} configurations and \gls{NCR} gain.

\section{Conclusion}
This paper presented an optimized deployment framework for \gls{HSRE} networks, designed to minimize infrastructure costs while ensuring full coverage in urban environments. By integrating \gls{RIS} and \gls{NCR} devices and considering realistic environmental constraints, our approach effectively meets the high connectivity demands of dense urban areas. Through the proposed cost-minimization model, we demonstrated that balancing device configurations—such as \gls{RIS} dimensions and \gls{NCR} amplification gain—significantly impacts overall deployment costs, particularly under varying \gls{SNR} thresholds and connectivity requirements. In future work, we aim to explore an alternative optimization method to maximize coverage within a fixed budget, and to expand our framework to incorporate additional \gls{SRE} devices beyond \gls{RIS} and \gls{NCR}, further enhancing network planning flexibility.

\bibliographystyle{IEEEtran}
\bibliography{main.bib}
\end{document}